# Comparison of Credential Management Systems Based on the Standards of IEEE, ETSI, and YD/T 3957-2021

Abel C. H. Chen
*Information & Communications Security Laboratory,
Chunghwa Telecom Laboratorie*
ORCID: 0000-0003-3628-3033

*Abstract*—As V2X (Vehicle-to-Everything) technology becomes increasingly prevalent, the security of V2X networks has garnered growing attention worldwide. In North America, the IEEE 1609 series standards are primarily used, while Europe adopts the ETSI series standards, and China has also established its industry standard, YD/T 3957-2021, among others. Although these standards share some commonalities, they also exhibit differences. To achieve compatibility across these standards, analyzing their similarities and differences is a crucial issue. Therefore, this study focuses on analyzing the three major standards mentioned above, discussing aspects such as certificate formats, signed message formats, and certificate request processes. Additionally, this research evaluates the efficiency of different cryptographic methods, including NIST P-256 and SM2-256, SHA-256 and SM3-256, as well as AES-128 and SM4-128. Finally, the study implements these three major standards on V2X devices and compares the efficiency of message signing and signature verification in V2X systems, providing a reference for the development of a secure certificate management system for V2X networks.

## I. Introduction

In recent years, to enhance the security of V2X networks, various organizations around the world have begun developing standards for V2X certificate management systems tailored to their respective national conditions and environments. In North America, the primary standards are IEEE 1609.2 [1] and IEEE 1609.2.1 [2], along with the design of the Security Credential Management System (SCMS) [3]-[4]. In Europe, the main standards are ETSI TS 103 097 [5] and ETSI TS 102 941 [6], and the Cooperative-Intelligent Transport Systems (C-ITS) SCMS (CCMS) [7]-[8] has been developed. Meanwhile, China has currently established the industry standard YD/T 3957-2021 [9], among others, and has developed the Chinese SCMS (C-SCMS) [10].

Although different organizations have established distinct standards, efforts have been made to ensure compatibility between these standards to enable communication and authentication across V2X devices. As a result, recent international standards have gradually adopted more unified formats, while allowing parameter values to be set according to specific national requirements. In light of this, this study focuses on analyzing the similarities and differences between the three major standards, discussing their strengths and weaknesses, and evaluating and verifying their performance. The contributions of this research are outlined as follows:

- This study compares the V2X security standards adopted by different countries, including the IEEE 1609 series standards, the ETSI series standards, and the YD/T 3957-2021 and related standards.
- This study implements and compares the computational efficiency of various cryptographic standards.
- This study implements and compares the computational efficiency of different V2X security standards.

This paper is divided into eight sections. Section II and Section III analyze the structure of certificates and signed messages. Section IV and Section V analyze the certificate request process. Section VI compares the efficiency of different cryptographic methods, while Section VII compares the efficiency of different standards on V2X devices. Finally, Section VIII summarizes the findings of this study.

## II. The Structure of Certificates

To achieve compatibility between certificates from different standards, in recent years, various standards have begun to adopt the CertificateBase format defined by the IEEE 1609.2 standard (as shown in Fig. 1) [1]. Based on the issuer, certificates can be divided into two main categories: enrollment certificates and authorization certificates, which are described as follows.

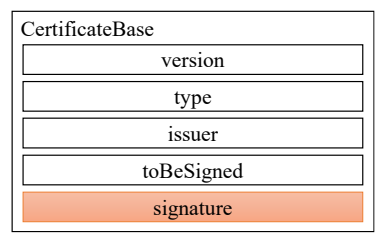

Fig. 1. The structure of CertificateBase.

### A. Enrollment Certificate / Enrolment Certificate

In both the SCMS and C-SCMS, each V2X device (referred to as an End Entity (EE)) can be issued an enrollment certificate by the Enrollment Certificate Authority (ECA), which serves as the legal identification for the V2X device. Similarly, in the CCMS, each V2X device (referred to as an ITS Station (ITS-S)) can be issued an enrolment certificate by the Enrolment Authority (EA) to serve as its legal identification. Both the enrollment certificate and the enrolment certificate are based on the CertificateBase format.

## B. Authorization Certificate / Authorization Ticket

In both the SCMS and C-SCMS, each V2X device can use the issued enrollment certificate to request an authorization certificate from the Authorization Certificate Authority (ACA), which is used when the V2X device sends signed messages. Similarly, in the CCMS, each V2X device can use the issued enrolment certificate to request an authorization ticket from the Authorization Authority (AA) for use when sending signed messages. Both the authorization certificate and the authorization ticket are based on the CertificateBase format.

## C. Note

It is worth noting that in the type field, SCMS defaults to using implicit certificates, while CCMS and C-SCMS default to using explicit certificates. Additionally, although the 2022 version of IEEE 1609.2 has added support for SM2, SM3, and SM4, the cryptographic methods or parameters adopted by different standards still vary slightly, as shown in Table I. For example, in SCMS, signing is performed using the Elliptic Curve Digital Signature Algorithm (ECDSA) [11] with NIST P-256; in CCMS, signing is done using the ECDSA with Brainpool P-256; and in C-SCMS, signing uses the SM2 signature algorithm [12] with SM2-256.

TABLE I. DEFAULT CRYPTOGRAPHY METHODS

| Cryptography | SCMS | CCMS | C-SCMS |
|---|---|---|---|
| Asymmetric Cryptography | NIST P-256 | Brainpool P-256 | SM2-256 |
| Hash | SHA-256 | SHA-256 | SM3-256 |
| Symmetric Cryptography | AES-128 | AES-128 | SM4-128 |

## III. THE STRUCTURE OF SIGNED MESSAGES

To achieve compatibility among V2X packets across different standards, various standards have begun to adopt the Ieee1609Dot2Data format in recent years. Among these, the most commonly used packets are signed messages, primarily defined in Ieee1609Dot2Data-Signed (as shown in Fig. 2) [2], and signed and encrypted messages, mainly defined in Ieee1609Dot2Data-SignedEncrypted (as shown in Fig. 3) [2].

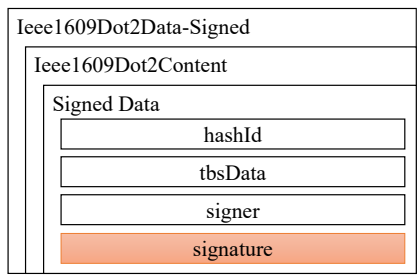

Fig. 2. The structure of Ieee1609Dot2Data-Signed.

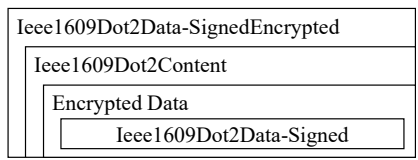

Fig. 3. The structure of Ieee1609Dot2Data-SignedEncrypted.

## A. Ieee1609Dot2Data-Signed in the IEEE's SCMS

In SCMS, the structure of Ieee1609Dot2Data-Signed [2] is shown in Fig. 2 and mainly includes four fields: hashId, tbsData, signer, and signature. Using the example of sending a basic safety message (BSM), the EE sets the hashId to SHA-256, the tbsData to the ITS application message, the signer to the hash value of the authorization certificate, and generates the signature using ECDSA with NIST P-256, placing the resulting signature in the signature field.

## B. EtsiTs103097Data-Signed in the ETSI's CCMS

In CCMS, the structure of EtsiTs103097Data-Signed [5] is inherited from the structure of Ieee1609Dot2Data-Signed, with the field information being the same as shown in Fig. 2. It is worth noting that ETSI has separately defined Cooperative Awareness Message (CAM) and Decentralized Environmental Notification Message (DENM), which inherit from EtsiTs103097Data-Signed, and thus still follow the Ieee1609Dot2Data-Signed format. Using the example of sending a CAM, the ITS-S sets the hashId to SHA-256, the tbsData to the ITS application message, the signer to the hash value of the authorization ticket, and generates the signature using ECDSA with Brainpool P-256, placing the resulting signature in the signature field.

## C. V2XSecData-Signed in the C-SCMS

In C-SCMS, the structure of V2XSecData-Signed [9] is inherited from the structure of Ieee1609Dot2Data-Signed, with the field information being the same as shown in Fig. 2, primarily including the four fields: hashId, tbsData, signer, and signature. In terms of naming, "Ieee1609Dot2" is replaced with "V2XSec," and Provider Service Identifier (PSID) is changed to Application Identifier (AID), while most other structures are inherited from the IEEE standard. Using the example of sending a Basic Safety Message (BSM), the EE sets the hashId to SHA-256, the tbsData to the ITS application message, the signer to the hash value of the authorization certificate, and generates the signature using the SM2 signature algorithm with SM2-256, placing the resulting signature in the signature field.

## IV. THE COMPARISON OF ENROLLMENT CERTIFICATE REQUESTS

## A. The Enrollment Certificate Request in the IEEE's SCMS

In SCMS, the process starts with the EE sending an EeEcaCertRequestSpdu to the ECA, signing the request using the pre-loaded canonical private key, and including the enrollment public key in the request. The EeEcaCertRequestSpdu is a type of signed certificate request. After the ECA verifies the signature and the content of the request using the pre-loaded canonical public key, the ECA sends an EcaEeCertResponseSpdu to the EE, which contains the EE's enrollment certificate. The ECA signs this response, and the EcaEeCertResponseSpdu is a form of Ieee1609Dot2Data-Signed, as shown in Fig. 4.

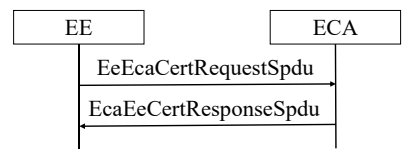

Fig. 4. The enrollment certificate request in the IEEE's SCMS.

## B. The Enrolment Certificate Request in the ETSI's CCMS

In CCMS, the process begins with the EE sending an EnrolmentRequest to the EA, signing the request using both the pre-loaded canonical private key and the enrolment private key. The request includes the enrollment public key and is encrypted using the EA's public key. The EnrolmentRequest is a type of Ieee1609Dot2Data-SignedEncrypted. After the EA verifies the signature and the content of the request using the pre-loaded canonical public key and the enrolment public key from the request, the EA sends an EnrolmentResponse to the EE. The EnrolmentResponse includes the EE's enrolment certificate and is signed and encrypted by the EA. The EnrolmentResponse is a form of Ieee1609Dot2Data-SignedEncrypted, as shown in Fig. 5.

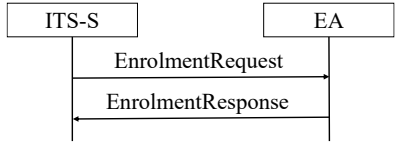

Fig. 5. The enrolment certificate request in the ETSI's CCMS.

## C. The Enrollment Certificate Request in the C-SCMS

In C-SCMS, secure connections and certificate requests can be established using the Device Configuration Manager (DCM) mechanism and the Generic Bootstrapping Architecture (GBA) mechanism. Taking the GBA mechanism defined by 3GPP [13] as an example, the EE establishes a secure connection with the GBA authentication system (AS), after which it sends the EeEcaCertRequestSpdu and receives the EcaEeCertResponseSpdu over this secure connection, as shown in Fig. 6. The EeEcaCertRequestSpdu and EcaEeCertResponseSpdu are consistent with the description in Subsection IV.A.

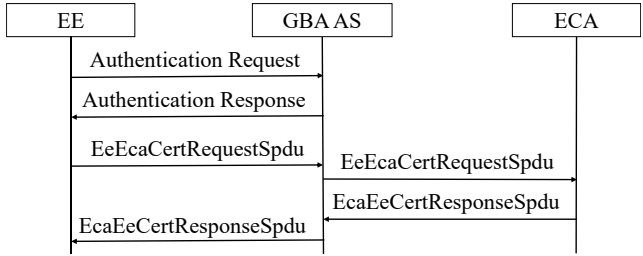

Fig. 6. The enrollment certificate request in the C-SCMS.

## V. THE COMPARISON OF AUTHORIZATION CERTIFICATE REQUESTS

### A. The Authorization Certificate Request in the IEEE's SCMS

In SCMS, after the EE obtains the enrollment certificate, it can send an EeRaCertRequestSpdu to the Registration Authority (RA), signing the request with the enrollment private key. This request includes the EE's enrollment certificate and the caterpillar public key, and is then encrypted using the RA's public key. The EeRaCertRequestSpdu is a type of signed and encrypted message. The RA retrieves the EE's enrollment public key from the enrollment certificate to verify the signature and the content of the request. After successful verification, the RA sends an RaEeCertAckSpdu to the EE. Additionally, the RA generates a cocoon public key based on the caterpillar public key and sends an RaAcaCertRequestSpdu to the ACA, signing the request with the RA's private key, which includes the cocoon public key. The RaAcaCertRequestSpdu is a type of signed message. After the ACA verifies the signature and the request content, it sends an AcaRaCertResponseSpdu to the RA. The AcaRaCertResponseSpdu contains the AcaResponse, which includes the EE's authorization certificate and the authorization certificate containing the butterfly public key (or the reconstruction value of the implicit certificate). Finally, the EE sends an EeRaDownloadRequestSpdu to the RA, requesting to download the authorization certificate. The RA provides a zip file containing RaEeCertInfo and AcaResponse to the EE, as shown in Fig. 7. The EeRaDownloadRequestSpdu is a type of Ieee1609Dot2Data-SignedEncrypted. It is worth noting that in SCMS, the RA is responsible for verifying the legitimacy of the EE's enrollment certificate.

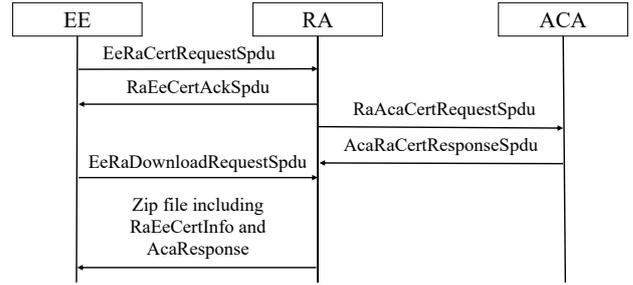

Fig. 7. The authorization certificate request in the IEEE's SCMS.

### B. The Authorization Ticket Request in the ETSI's CCMS

In CCMS, after the ITS-S obtains the enrolment certificate, it can send an AuthorizationRequest to the AA, signing the request with the authorization private key. This request includes the authorization public key and EcSignature, where the EcSignature is signed by the EE using the enrolment certificate and enrolment private key. The request is then encrypted using the AA's public key. The AuthorizationRequest is a type of signed and encrypted message. It is important to note that at this point, the AA only uses the authorization public key to verify the signature of the request and does not verify the legitimacy of the ITS-S's enrolment certificate. The EcSignature is encapsulated in the AuthorizationValidationRequest, which is signed by the AA using its private key and encrypted using the EA's public key before being sent to the EA. When the EA receives the AuthorizationValidationRequest, it verifies both the signature of the AuthorizationValidationRequest and the signature of the EcSignature, and confirms the related information. Upon successful verification, the EA replies with an AuthorizationValidationResponse to the AA. Both the AuthorizationValidationRequest and AuthorizationValidationResponse are types of signed and encrypted messages. Finally, the AA issues the ITS-S's authorization ticket and encapsulates it in the AuthorizationResponse, which is sent back to the ITS-S, as shown in Fig. 8. It is worth mentioning that there is no RA role in CCMS, and the EA in CCMS takes over some of the functions of the RA in SCMS, such as verifying the legitimacy of the ITS-S's enrolment certificate.

### C. The Authorization Certificate Request in the C-SCMS

In C-SCMS, the DCM mechanism and GBA mechanism can also be used to establish a secure connection and apply for the authorization certificate request. Taking the GBA mechanism defined by 3GPP [13] as an example, the EE establishes a secure connection with the GBA AS through

Authentication Request and Authentication Response. After establishing this secure connection, the EE sends an EeRaCertRequestSpdu and receives a RaEeCertAckSpdu, as shown in Fig. 9. The process involving EeRaCertRequestSpdu, RaEeCertAckSpdu, and the subsequent steps are consistent with those described in Subsection V.A. It is worth noting that in C-SCMS, the RA is referred to as the Pseudonymous Registration Authority (PRA).

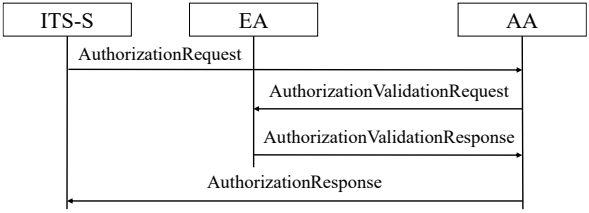

Fig. 8. The authorization ticket request in the ETSI's CCMS.

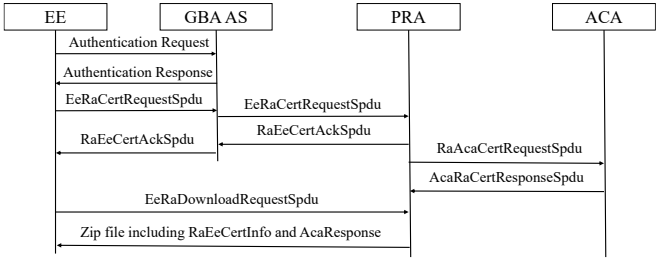

Fig. 9. The authorization certificate request in the C-SCMS.

## VI. THE PERFORMANCE COMPARSION OF CRYPTOGRAPHY METHODS

To evaluate the computational efficiency of different cryptographic methods, this section implements the cryptographic methods listed in Table I on a laptop. The hardware and software specifications include an Intel(R) Core(TM) i7-10510U CPU @ 1.80GHz, 16.0 GB RAM, Windows 10 Enterprise Edition, Java 18.0.2.1, and BouncyCastle 1.78. The following subsections will separately assess the performance of elliptic curve cryptography (ECC) digital signatures, ECC encryption and decryption, hashing algorithms, and symmetric encryption and decryption.

### A. NIST P-256 and SM2-256 For Signing

This subsection evaluates the computational efficiency of the ECDSA and SM2 signature algorithms. ECDSA [11] is paired with the elliptic curve NIST P-256, and the SM2 signature algorithm [12] is paired with the elliptic curve SM2-256. Each algorithm was executed 3000 times to measure the key generation time, signature generation time, and signature verification time. The experimental results are shown in Fig. 10.

From the results, it can be observed that the key generation process is similar for both ECDSA and the SM2 signature algorithm, as both involve randomly generating an integer and then computing an elliptic curve point based on that value, leading to no significant difference in computation time. However, in the signature generation process, SM2 requires an additional hashing calculation, which results in more computation time for signature creation. Finally, for signature verification, there is no significant difference between ECDSA and the SM2 signature algorithm.

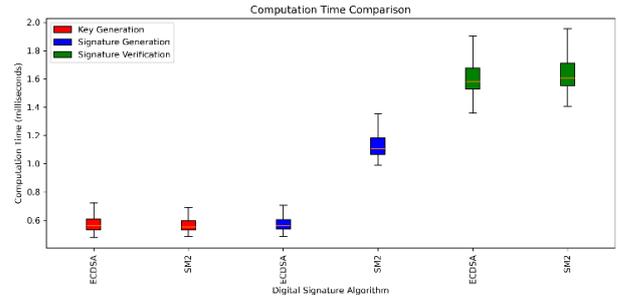

Fig. 10. The comparison of NIST P-256 and SM2-256 for signing.

### B. NIST P-256 and SM2-256 For Encryption

This subsection evaluates the computational efficiency of the ECIES (Elliptic Curve Integrated Encryption Scheme), commonly used for encryption and decryption, which primarily facilitates the encapsulation or negotiation of a shared symmetric key. The algorithms tested include ECIES with elliptic curve NIST P-256 [14] and ECIES with elliptic curve SM2-256 [15], each executed 3,000 times to measure key generation time, key encapsulation time, and key extraction time. The experimental results are shown in Fig. 11.

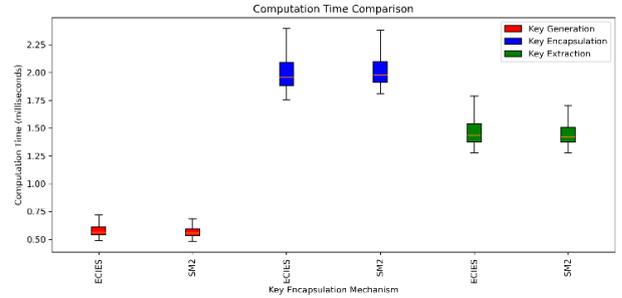

Fig. 11. The comparison of NIST P-256 and SM2-256 for encryption.

From the results, it can be observed that under the same ECIES algorithm, whether using elliptic curve NIST P-256 or elliptic curve SM2-256, there is no significant difference in key generation time, key encapsulation time, or key extraction time.

### C. SHA-256 and SM3-256

This subsection evaluates the computational efficiency of different hashing algorithms, specifically implementing and testing SHA-256 [16] and SM3-256 [15]. In the experiment, 3,000 hashing operations were executed to measure the hashing computation time. The experimental results are shown in Fig. 12.

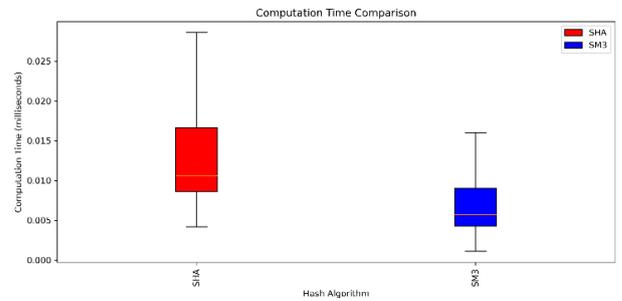

Fig. 12. The comparison of SHA-256 and SM3-256.

From the results, it can be observed that SM3-256 has higher computational efficiency, with a significantly lower computation time compared to SHA-256. Therefore, SM3-256 could be considered in the future to improve computational efficiency.

*D. AES-128 and SM4-128*

This subsection evaluates the computational efficiency of different symmetric encryption and decryption algorithms by implementing and testing AES-128 [17] and SM4-128 [18]. In the experiment, each algorithm was executed 3,000 times to measure both encryption and decryption computation times. The experimental results are shown in Fig. 13.

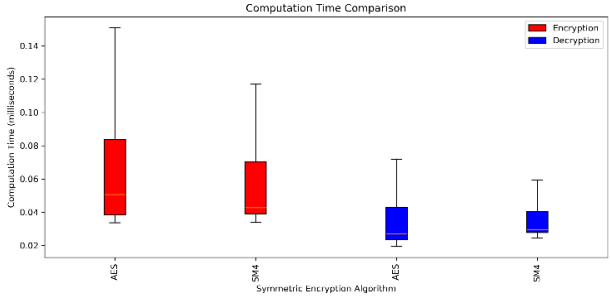

Fig. 13. The comparison of AES-128 and SM4-128.

From the results, it can be observed that SM4-128 has slightly lower computation times for both encryption and decryption compared to AES-128. Therefore, while SM4-128 demonstrates higher computational efficiency, the difference is not statistically significant, indicating that the computational efficiencies of AES-128 and SM4-128 are similar.

## VII. IMPLEMENTATION IN V2X DEVICES

To verify the execution performance of the SCMS, CCMS, and C-SCMS standards on V2X devices, this section will implement and test them using trusted electronic devices. Subsection VII.A introduces the experimental environment, while Subsection VII.B presents the comparative experimental results of the computation times.

*A. Experimental Enviroment*

In this study's experimental environment, Clientron's On-Board Units (OBUs) are used as V2X devices to implement the SCMS, CCMS, and C-SCMS standards. The hardware and software specifications of the V2X devices include the MT2712 chip, Android 10, and BouncyCastle 1.78. After each V2X device applies for its enrollment certificate/enrolment certificate and authorization certificate/authorization ticket, they send and receive BSMs according to the Ieee1609Dot2Data-Signed format defined in IEEE 1609.2.1 [2]. The implementation screen is shown in Fig. 14.

*B. Computation Time Comparison*

When sending a BSM, the V2X device needs to encapsulate the BSM into the Ieee1609Dot2Data-Signed format and generate a signature for the message to prove the legitimacy of the sender. The authorization certificate/authorization ticket is placed in the signer field to provide verification for the recipient. Fig. 15 shows the computation time required to generate an Ieee1609Dot2Data-Signed message containing BSM information under the SCMS, CCMS, and C-SCMS standards. It can be observed that since both SCMS and CCMS use ECDSA, the computation time does not show significant differences, even though they utilize different elliptic curves, NIST P-256 and Brainpool P-256, respectively. However, C-SCMS uses the SM2 signature algorithm, which, as described in Subsection VI.A, requires more computation time. Therefore, the C-SCMS implementation takes longer to produce packets containing BSM information.

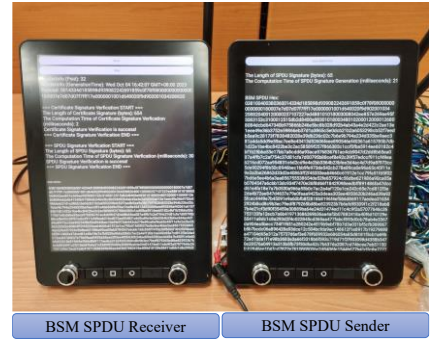

Fig. 14. The implementation of V2X devices based on Clientron's OBUs.

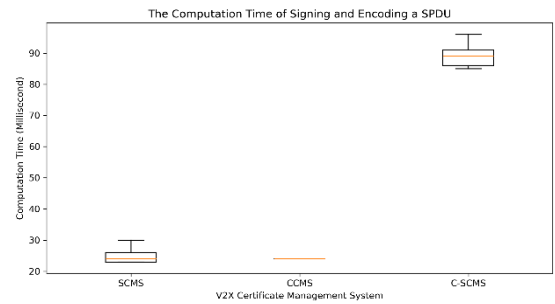

Fig. 15. The computation time comparison of signature generation.

When verifying packets containing a BSM, the receiving V2X device needs to obtain the authorization certificate/authorization ticket from the packet. It then uses the authorization certificate/authorization ticket to retrieve the public key of the sending V2X device to verify the packet. Fig. 16 shows the computation time required to verify an Ieee1609Dot2Data-Signed message with BSM information under the SCMS, CCMS, and C-SCMS standards. It can be observed that both CCMS and C-SCMS use explicit certificates by default, so the certificates contain the actual public key of the sending V2X device. This allows for direct verification of the signature using the public key, resulting in shorter computation times. In contrast, SCMS uses implicit certificates by default, which only include the reconstruction value. Therefore, it requires the ACA public key to reconstruct the actual public key of the sending V2X device, leading to longer computation times.

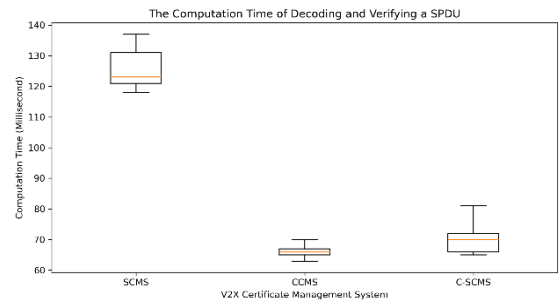

Fig. 16. The computation time comparison of signature verification.

## VIII. CONCLUSIONS AND FUTURE WORK

This study primarily explores the three standards: SCMS, CCMS, and C-SCMS, comparing and validating them based on differences in message flow and cryptographic methods, with a comprehensive discussion and verification from principles to implementation.

The research findings indicate that international standards are gradually becoming compatible. The certificate formats are now fully compatible, and the signed message packet formats can also align. The main differences lie in the certificate application processes and the packets used for certificate requests, as well as the cryptographic methods and their parameters. Furthermore, this study found that ECDSA outperforms SM2 in signing efficiency, while SM3 demonstrates better performance than SHA in hash computation efficiency. Additionally, CCMS and C-SCMS use explicit certificates by default, which results in higher efficiency in verifying secure packets compared to SCMS.

Although this study found that SCMS has lower efficiency in verifying secure packets, this may be an unfair comparison. Future work could implement implicit certificates in CCMS and C-SCMS, allowing for a comparison of all three standards based on implicit certificates.

## ACKNOWLEDGMENT

The RCA certificate developed in this study has successfully undergone interoperability testing at the OmniAir Plugfest conference and has been officially listed in the Test Certificate Trust List (CTL) by the SCMS Manager, available at: https://www.scmsmanager.org/publications/. Additionally, this certificate has also been recognized and included in the European Certificate Trust List (ECTL), maintained by the EU C-ITS Point of Contact (CPOC), which can be accessed at: https://cpoc.jrc.ec.europa.eu/ECTL.html. We are deeply grateful to OmniAir, the SCMS Manager, CPOC, and all other partners for their crucial support and contributions throughout this process.

## REFERENCES

[1] "IEEE Approved Draft Standard for Wireless Access in Vehicular Environments--Security Services for Applications and Management Messages," in *IEEE Std 1609.2-2022 (Revision of IEEE Std 1609.2-2016)*, pp. 1-349, March 2023, doi: 10.1109/IEEESTD.2023.10075082.

[2] "IEEE Standard for Wireless Access in Vehicular Environments (WAVE) - Certificate Management Interfaces for End Entities," in *IEEE Std 1609.2.1-2022 (Revision of IEEE Std 1609.2.1-2020)*, pp. 1-261, June 2022, doi: 10.1109/IEEESTD.2022.9810154.

[3] B. Brecht et al., "A Security Credential Management System for V2X Communications," in *IEEE Transactions on Intelligent Transportation Systems*, vol. 19, no. 12, pp. 3850-3871, Dec. 2018, doi: 10.1109/TITS.2018.2797529.

[4] A. C. H. Chen, B. -Y. Lin, C. -K. Liu and C. -F. Lin, "Implementation and Performance Analysis of Security Credential Management System Based on IEEE 1609.2 and 1609.2.1 Standards," *2023 IEEE International Conference on Machine Learning and Applied Network Technologies (ICMLANT)*, San Salvador, El Salvador, 2023, pp. 1-5, doi: 10.1109/ICMLANT59547.2023.10372990.

[5] "Intelligent Transport Systems (ITS); Security; Security header and certificate formats; Release 2," in *ETSI TS 103 097 V2.1.1*, pp. 1-22, October 2021.

[6] "Intelligent Transport Systems (ITS); Security; Trust and Privacy Management; Release 2," in *ETSI TS 102 941 V2.2.1*, pp. 1-94, November 2022.

[7] Y. Gong and B. -J. Hu, "A Quantum-Resistant Key Management Scheme Using Blockchain in C-V2X," in *IEEE Transactions on Intelligent Transportation Systems*, early access, doi: 10.1109/TITS.2024.3421381.

[8] A. C. H. Chen, B. -Y. Lin, C. -K. Liu and C. -F. Lin, "V2X Credential Management System Comparison Based on IEEE 1609.2.1 and ETSI TS 102 941," 2024 IEEE North Karnataka Subsection Flagship International Conference (NKCon), Belagavi, India, 2023, pp. 1-6.

[9] "LTE-based Vehicular Communication--Technical Requirement of Security Certificate Management System," in *Ministry of Industry and Information Technology Announcement*, *YD/T 3957-2021*, pp. 1-173, Dec. 2021.

[10] X. Wu, H. Ning, S. Sun, K. Qian, J. Fang, X. Wang and B. Wang, " Suggestions on Management of Electronic Authentication Services in Internet of Vehicles," in *Information and Communications Technology and Policy*, vol. 50, no. 3, pp. 60-65, March 2024, doi: 10.12267/j.issn.2096-5931.2024.03.009.

[11] X. Yang, M. Liu, M. H. Au, X. Luo and Q. Ye, "Efficient Verifiably Encrypted ECDSA-Like Signatures and Their Applications," in *IEEE Transactions on Information Forensics and Security*, vol. 17, pp. 1573-1582, 2022, doi: 10.1109/TIFS.2022.3165978.

[12] Y. Zhang et al., "Verifiable Random Function Schemes Based on SM2 Digital Signature Algorithm and its Applications for Committee Elections," in *IEEE Open Journal of the Computer Society*, vol. 5, pp. 480-490, 2024, doi: 10.1109/OJCS.2024.3463649.

[13] "3rd Generation Partnership Project; Technical Specification Group Services and System Aspects; Generic Authentication Architecture (GAA); Generic Bootstrapping Architecture (GBA)(Release 18)," in *3GPP TS 33.220 V18.3.0*, pp. 1-110, March 2024.

[14] P. Realpe-Muñoz, J. Velasco-Medina and G. Adolfo-David, "Design of an S-ECIES Cryptoprocessor Using Gaussian Normal Bases Over GF(2m)," in *IEEE Transactions on Very Large Scale Integration (VLSI) Systems*, vol. 29, no. 4, pp. 657-666, April 2021, doi: 10.1109/TVLSI.2021.3057985.

[15] X. Zheng, C. Xu, X. Hu, Y. Zhang and X. Xiong, "The Software/Hardware Co-Design and Implementation of SM2/3/4 Encryption/Decryption and Digital Signature System," in *IEEE Transactions on Computer-Aided Design of Integrated Circuits and Systems*, vol. 39, no. 10, pp. 2055-2066, Oct. 2020, doi: 10.1109/TCAD.2019.2939330.

[16] A. C. H. Chen, "Report on Hash Algorithm Performance — A Case Study of Cryptocurrency Exchanges Based on Blockchain System," *2024 IEEE International Conference on Data Intelligence and Cognitive Informatics (ICDICI)*, Tirunelveli, India, 2024, pp. 1-6.

[17] W. -K. Lee, H. J. Seo, S. C. Seo and S. O. Hwang, "Efficient Implementation of AES-CTR and AES-ECB on GPUs With Applications for High-Speed FrodoKEM and Exhaustive Key Search," in *IEEE Transactions on Circuits and Systems II: Express Briefs*, vol. 69, no. 6, pp. 2962-2966, June 2022, doi: 10.1109/TCSII.2022.3164089.

[18] W. Liu et al., "A 128-Gbps Pipelined SM4 Circuit With Dual DPA Attack Countermeasures," in *IEEE Transactions on Very Large Scale Integration (VLSI) Systems*, vol. 32, no. 6, pp. 1164-1168, June 2024, doi: 10.1109/TVLSI.2024.3379205.